\documentclass[twocolumn,showpacs,preprintnumbers,amsmath,amssymb]{revtex4}

\usepackage{graphicx}
\usepackage{dcolumn}
\usepackage{bm}
\usepackage[left]{lineno}
\usepackage[usenames]{color}

\begin{document}

\preprint{APS/123-QED}

\title{

Ballistic-Ohmic quantum Hall plateau transition in graphene pn junction}

\author{Tony Low}
\email{tonyaslow@gmail.com}
\affiliation{School of Electrical \& Computer Engineering, Purdue University, West Lafayette, IN47906, USA }%

\date{\today}

\begin{abstract}
Recent quantum Hall experiments conducted on disordered graphene pn junction provide evidence that the junction resistance could be described by a simple Ohmic sum of the n and p mediums' resistances. However in the ballistic limit, theory predicts the existence of chirality-dependent quantum Hall plateaus in a pn junction. We show that two distinctively separate processes are required for this ballistic-Ohmic plateau transition, namely (i) hole/electron Landau states mixing and (ii) valley isospin dilution of the incident Landau edge state. These conclusions are obtained by a simple scattering theory argument, and confirmed numerically by performing ensembles of quantum magneto-transport calculations on a $0.1\mu m$-wide disordered graphene pn junction within the tight-binding model. The former process is achieved by pn interface roughness, where a pn interface disorder with a root-mean-square roughness of $10nm$ was found to suffice under typical experimental conditions. The latter process is mediated by extrinsic edge roughness for an armchair edge ribbon and by intrinsic localized intervalley scattering centers at the edge of the pn interface for a zigzag ribbon. In light of these results, we also examine why higher Ohmic type plateaus are less likely to be observable in experiments.

\end{abstract}

\maketitle

\section{\label{sec:level1}INTRODUCTION}

Graphene is a two-dimensional sheet of carbon atoms arranged in a honeycomb lattice with unique electronic properties i.e. linear energy dispersion with zero bandgap described by the relativistic Dirac equation \cite{novoselov04,semenoff84}. This attribute manifests itself as anomaly in the quantum Hall regime \cite{novoselov05,zhang05}, where the Landau filling factor, $\nu$, goes by the non-conventional sequence of $4n+2$ and $n=0,1,2,\ldots$ labels the Landau levels (LL). This leads to conductance plateaus at $\sigma=\nu$, expressed in units of the von Klitzing constant $e^2/h$. Room temperature observations \cite{zhang05} of these quantum Hall plateaus make it a potential state variable for possible future device applications, such as quantum computation using Landau levels as the basic qubit \cite{beenakker03} and pure spin current switching devices \cite{abanin06}.  

Recently, experiments by Williams $et$ $al.$ \cite{williams07} on the magneto-transport of graphene pn junction in the quantum Hall regime found that the two terminal junction conductance ($\sigma_{pn}$) exhibits new plateaus. These conductance plateaus were predicted to follow a simple Ohmic conductance rule, first proposed by Abanin $et$ $al.$ \cite{abanin07}, 
\begin{eqnarray}
\hat{\sigma}_{pn}(\nu_{n},\nu_{p})=\left(\frac{1}{\nu_{n}}+\frac{1}{\nu_{p}}\right)^{-1}=\left(\frac{\nu_{p}}{\nu_{n}+\nu_{p}}\right)_{eq}\nu_{n}
\label{ohmic}
\end{eqnarray}
where $\nu_{n/p}$ are the Landau levels filling factors in the n/p regions. The last expression in Eq. \ref{ohmic} presents the physics more lucidly, with $(\ldots)_{eq}$ embodying the Landau modes mixing process along the pn interface for the electron current. Experiments \cite{williams07}, however, did not reproduce all the predicted plateaus, especially for the higher filling factor combination values i.e. such as $(\nu_{n},\nu_{p})=(2,6)$ and $(6,6)$\footnote{A more recent experiment by Lohmann $et$ $al.$ \cite{lohmann09}, which employed a four terminal measurement, observed prominent plateaus for only $(2,2)$. The higher filling factor combination values were not reported. In their experiments, the pn junctions were created via chemical doping, instead of the usual top/bottom gating scheme \cite{williams07}. }. This observation leads us in formulating our first question, ``what degree of interface disorder is required to observe complete modes mixing, especially for the cases with higher combination values?''. 

Interestingly, the appearance of quantum Hall plateaus in a graphene pn junction is not isolated to only disordered samples. Tworzydlo and co-workers \cite{tworzydlo07} showed theoretically that new plateaus should be observed in perfectly clean graphene ribbons with perfect edges. These plateaus are independent of $(\nu_{n},\nu_{p})$, and depend only on the ribbons' chirality. We herein denote these ballistic plateaus as $\tilde{\sigma}_{pn}$, and they are given by \cite{tworzydlo07},
\begin{eqnarray}
\tilde{\sigma}_{pn}=\left\{
\begin{array}{cc}
\tfrac{1}{2} \mbox{ or } 2 & \mbox{armchair}\\
0 \mbox{ or } 2 & \mbox{zigzag}
\end{array}
\right.
\label{clean}
\end{eqnarray}
where the widths dictate the above possible outcomes. The underlying physics for this width dependence is explained in terms of the valley isospin (for armchair-edge \cite{tworzydlo07}) and parity (for zigzag-edge \cite{akhmerov08}) of the lowest lying Landau modes.  The above revelation begs the  second question, ``Is pn interface disorder alone sufficient in inducing the plateau transition from $\tilde{\sigma}_{pn}$ to $\hat{\sigma}_{pn}$, or is edge disorder (or other intervalley scattering processes) also necessary?''. 

Theoretical studies on the effect of disorder on the magneto-transport properties of a graphene pn junction are few \cite{tworzydlo07,long08,li08}. In the classical integer quantum Hall problem, the tight-binding Hamiltonian model \cite{anderson58} and the Chalker/Coddington's network model \cite{chalker88} are popular approaches for studying magneto-transport in disordered system \cite{huckestein95}. For this problem the former is more suitable, since the formalism inherently captures both quantum mechanical and atomistic effects which are believed to play an important role. Essentially, one seeks the direct solution to the one-electron Schr$\ddot{o}$dinger equation, where the open boundary scattering problem \cite{frensley90} is usually conceptualized within the Landauer-B\"{u}ttiker quantum transmission point-of-view \cite{landauer70,buttiker86}. The quantum transmission function can be calculated using the Green's function \cite{datta97,haug96} or wave function approach \cite{lent90}, the former being a more popular technique in recent years. Within this theoretical framework, Long and co-workers \cite{long08} conducted an intensive statistical study of disordered graphene pn junctions in the quantum Hall regime. Graphene ribbons with zigzag edges are considered in their work. Bulk disorder was incorporated into their Hamiltonian in the form of on-site energy fluctuations. They found that the Ohmic type quantum Hall plateaus, $\hat{\sigma}_{pn}$, emerges with sufficient disorder strength. Similar conclusions had also been reached by Li and Shen \cite{li08}, where pn interface disorder was considered for a zigzag ribbon, also in the form of on-site energy fluctuations. Both these numerical studies \cite{long08,li08} employed short range disorder potential which varies on the scale of the lattice constant. Although the actual sources of disorder varies across different experimental samples \cite{neto09}, disorder such as pn interface and edge roughness are not justifiably captured by short range disorder potential. Moreover, short range disorder potential serve to masked the effect of valley isospin, which is inevitably washed out by such disorder \cite{li08}. For example, Tworzydlo $et$ $al.$ \cite{tworzydlo07} had employed a long range disorder potential, and found that the junction magneto-conductance of a zigzag ribbon is extremely sensitive to the disorder, while an armchair ribbon shows the opposite behavior. 

The purpose of this paper is to quantitatively model the effect of pn interface and edge disorders and to establish a coherent, conceptual framework for understanding the plateau evolution from $\tilde{\sigma}_{pn}$ to $\hat{\sigma}_{pn}$. We reason that one can distinguish two underlying mechanisms which are prerequisites for this ballistic-Ohmic plateau transition, namely (i) hole and electron Landau states mixing and (ii) valley isospin dilution. The former can be achieved via pn interface roughness, a long range potential type disorder. The latter process is mediated by extrinsic edge roughness for an armchair edge ribbon and by intrinsic localized intervalley scattering centers at the edge of the pn interface for a zigzag ribbon. By performing ensembles of quantum magneto-transport calculations on a $0.1\mu m$-wide graphene pn junction with pn interface (ID) and edge disorder (ED), we illustrate how the plateaus evolve from $\tilde{\sigma}_{pn}$ to $\hat{\sigma}_{pn}$ for both armchair and zigzag ribbons.



This paper is organized as follows. Section II begins with an introduction of our quantum transport model based on the Green's function and tight-binding approaches. Computational and numerical aspects are highlighted in this section, including the statistical modeling of the interface and edge disorder. Section III discusses the magneto-transport across a pn junction in the ballistic limit. The underlying physics for the existence of the ballistic quantum Hall plateaus $\tilde{\sigma}_{pn}$ are reviewed and discussed. Section IV studies the ballistic-Ohmic quantum Hall plateau transition in the presence of ID/ED for both armchair and zigzag ribbons. An accompanying simple Chalker/Coddington type scattering theory is presented to elucidate the underlying physics. Section V pertains to the analysis and comparison with recent experimental data in the literatures followed by a summary of this work.

\begin{figure}[t]
\centering
\scalebox{0.5}[0.5]{\includegraphics*[viewport=180 230 620 520]{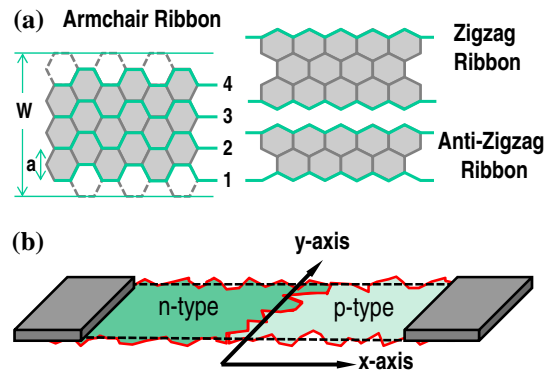}}
\caption{\footnotesize (a) Illustrations of armchair, zigzag and anti-zigzag edge ribbons. The carbon layer numbering convention for an armchair edge ribbon used in this work is also depicted. (b) Schematic of the two terminal pn junction simulated in this work, illustrating also the ID and ED.}
\label{IRorER}
\end{figure}

\section{\label{sec:level2} Quantum Transport Model}

In quantum Hall experiments, a magnetic field of the order of $10T$ is common. This corresponds to a magnetic length of $\ell_{B}\approx 20nm$, which approximates the spatial extent of the ground state LL wavefunction. The device's width has to be appropriately chosen so that the LL's wavefunctions along the opposite edges do not overlap, especially for the higher LLs.  However, finite computational resources set a practical limit on the matrix size of the system Hamiltonian. Based on the above considerations, we employed a $100nm$-wide graphene ribbon pn junction device, which is schematically shown in Fig.\ref{IRorER}c. Ribbons of both armchair and zigzag-type edges are considered in this work \cite{beenakker08,neto09}, as shown in Fig.\ref{IRorER}a. We seek to compute the junction conductance ($\sigma_{pn}$) in the presence of pn interface and edge disorder. For pedagogical purposes, we shall briefly review the quantum transport model employed in this work.

\subsection{\label{sec:2b} Green's function theory with a tight-binding model Hamiltonian}
In this work, the device is described by tight-binding Hamiltonian given by
\begin{eqnarray}
H=\sum_{i}v_{i}a_{i}^{\dagger}a_{i}+\sum_{ij}\left|t_{ij}\right|exp\left(i\frac{e}{\hbar}\int_{i}^{j} \bold{A}\cdot d\bold{l}\right)a_{i}^{\dagger}a_{j}, 
\label{tbmodel}
\end{eqnarray}
where $a_{i}^{\dagger}/a_{i}$ are the creation/destruction operators at each atomic site $i$, and $v_{i}$ and $t_{ij}$ are the on-site potential energy and hopping energies \cite{wallace47,saito98}. The simple one $p_z$-orbital description as described by Eq. \ref{tbmodel} is sufficient for the modeling of the relevant energy bands of interest for electronic transport properties \footnote{However, one could easily extends the simple tight binding Hamiltonian to include higher orbitals by extracting the orbital coupling energies from a density functional calculation as described in \cite{bevan09}.}. We assumed that $\left|t_{ij}\right|=3eV$. In the presence of a perpendicular magnetic field B, $t_{ij}$ will incorporate a Peierls phase $\phi_{ij}\left(\bold{A}\right)$, where $\bold{A}$ is the corresponding vector potential. The phase $\phi_{ij}$ is assigned such that the magnetic flux through an arbitrary area S satisfies the following Stokes law \cite{wakabayashi99},
\begin{eqnarray}
\frac{1}{\phi_{0}}\int B dS =\sum_{\Omega} \phi_{ij},
\label{stokeslaw}
\end{eqnarray}
where $\Omega$ is the boundary of S. The parameter $v_{i}$ is determined by the top/bottom gate electrostatics, which is known a priori.  The effect of pn interface and edge disorder are also described through $v_{i}$ and $t_{ij}$ respectively (discussions defered to next section).

The Landauer-B\"{u}ttiker approach pictures a device in which dissipative processes are absent but coupled to perfect thermodynamic systems known as ``reservoirs''. This approach has been very successful in modeling physical effects in a myraid of problems in the field of mesoscopic physics \cite{datta97,imry02,haug96,mahan90}. In numerical implementation, $H$ is divided into ``device'' ($H_{d}$) and ``contacts'' regions. $H_{d}$ is constructed so that it captures the scattering region of interests, hence it is of finite matrix size. The contacts represent the semi-infinite regions of $H$, which characterized the open boundary nature of the transport problem. Through simple algebras as detailed in \cite{datta97}, one could write a Green's function for the ``device'' as follows,
\begin{eqnarray}
G(\epsilon_{f})=\left(\epsilon_{f}-H_{d}-\Sigma_{s}-\Sigma_{d}\right)^{-1}
\label{green}
\end{eqnarray}
where $\epsilon_{f}$ is the Fermi energy. $\Sigma_{s/d}$ are conveniently known as the contact self-energies (subscript s/d for source/drain respectively), which could be expressed as $\Sigma_{s/d}=\tau_{s/d}g_{s/d}\tau_{s/d}^{\dagger}$, where $\tau_{s/d}$ describes the coupling of the s/d contacts to the device and $g_{s/d}$ is the surface Green's function of the respective contacts. In this work, $g_{s/d}$ is obtained using an efficient iterative scheme outlined in \cite{sancho84}. Direct matrix inversion of Eq. \ref{green} proves to be computationally prohibitive. Therefore, one commonly resorts to recursive type techniques, such as the recursive Green's function approach \cite{nonoyama98,anantram08}, the renormalization method \cite{grosso89}, or combination of both techniques \cite{low09}. After solving for $G(\epsilon_{f})$, we can compute the device conductance at $\epsilon_{f}$ via (in units of $\tfrac{e^{2}}{h}$)  \cite{datta97},
\begin{eqnarray}
\sigma_{pn}=2\mbox{Tr}\left(\Gamma_{s}G\Gamma_{d}G^{\dagger}\right),
\label{conduc}
\end{eqnarray}
where $\Gamma_{s/d}$ are known as the contact broadening functions which can be obtained from the respective self-energy i.e. $\Gamma_{s/d}=i(\Sigma_{s/d}-\Sigma_{s/d}^{\dagger})$. One can view Eq. \ref{conduc} as just a different form of the Fisher-Lee expression \cite{lee81}\cite{datta97}. Other physical observable quantities such as the local density of states (LDOS), charge density ($n(x,y)$) and current density ($\vec{j}(x,y)$) can also be obtained \cite{low09}.
 
\begin{figure}[t]
\centering
\scalebox{0.95}[0.95]{\includegraphics*[viewport=160 350 620 720]{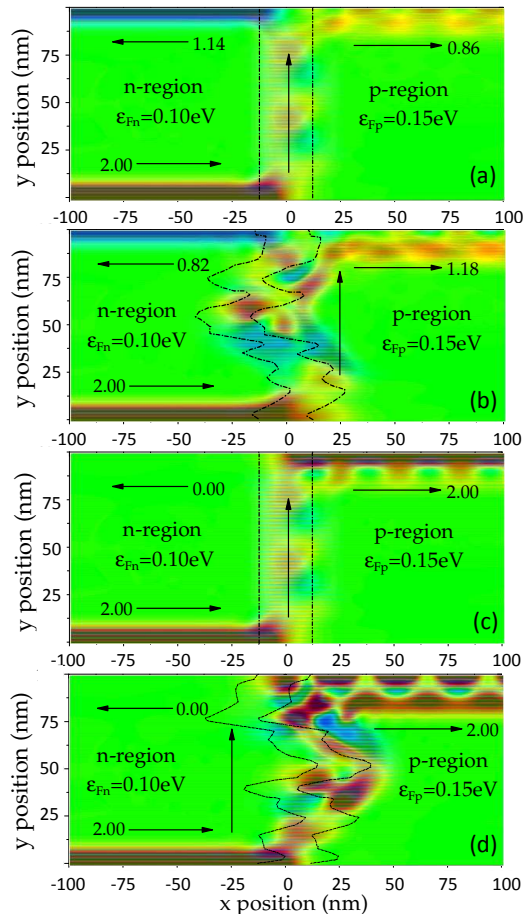}}
\caption{\footnotesize Intensity plot of the longitudinal current component for magneto-transport across an armchair graphene pn junction at $B=10T$. We consider ground state Landau level injected from the left. The biased condition is such that the filling factor combinations is $(\nu_{n},\nu_{p})=(2,6)$, where spin degeneracy accounted for. The depletion width is assumed to be $25nm$ and perfect edges is assumed. We plotted for the case $\bold{(a)}$ without ID and $\bold{(b)}$ with ID for ribbon with $401$ carbon layers along the width. Similary, for ribbon with $400$ carbon layers in $\bold{(c)}$ and $\bold{(d)}$.  }
\label{currentplots}
\end{figure}

\subsection{\label{sec:2c} Modeling interface/edge disorders}

In this work, we are interested in two types of disorder - interface and edge disorder (ID/ED). It is generally accepted that ID plays an important role in facilitating the mixing of the electron/hole Landau modes along the pn interface \cite{abanin07}. We define the pn interface as the equi-energy line of Dirac points separating the n/p regions. This interface will be highly susceptible to electrostatic influences of impurities due to ineffective charge screening in the depletion region. The presence of electron-hole puddles due to trapped impurities in the oxide layers \cite{hwang07} (or ripples \cite{herbut08}) was recently observed by Martin and co-workers \cite{martin08}. Charge density fluctuations of $\pm1\times 10^{11}cm^{-2}$ was reported in their work. This implies significant ID. Edge roughness was also recently characterized by Gupta $et$ $al.$ \cite{gupta09}, where a root-mean-square roughness of about $3nm$ was observed in micromechanically produced ribbons.

In order to model ID and ED, we devise a simple algorithm for generating a one dimensional roughness profile, $I(y)$. $I(y)$ can be expressed in its Fourier components,
\begin{eqnarray}
I(y)=\sum_{n}A_{n}sin\left(\tfrac{n\pi y}{W}\right)
\label{IR}
\end{eqnarray}
where $W$ is the device width, and $A_{n}$ is the amplitude for the sine components given by,
\begin{eqnarray}
A_{n}=R(D_{1})exp(-\tfrac{n}{D_{2}})
\label{an}
\end{eqnarray}
where $R(D_{1})$ outputs a uniformly distributed random number between $\pm D_{1}$. Eq. \ref{an} represents the power spectrum of the roughness morphology, however there is currently no characterization of ID/ED morphologies to justify such assumption. Nevertheless, it is known that an exponential power spectrum describes the surface morphology of Si/SiO$_{2}$ interfaces \cite{goodnick85,low04,low05,low08}, making it a natural guess for the ID/ED morphologies in our study. For a given set of disorder parameters $\left\{D_{1},D_{2}\right\}$, we compute the ensemble average of N samples to obtain the conductance $\sigma_{pn}$, where N is chosen to be from $100-200$ samples. The root-mean-square (RMS) and auto-correlation length (AL) of the interface roughness morphology are also computed. We defined AL to be the length at which the cross-correlation is $50\%$ of the auto-correlation. It can be shown that AL depends on $D_{2}$ and is relatively insensitive to $D_{1}$. A larger $D_{2}$ will yield more higher frequency components in $I(y)$, thereby decreasing the AL. 

In the presence of disorder, the device Hamiltonian's on-site and coupling energies have to be modified accordingly. For $v_{i}$, we have,
\begin{widetext}
\begin{eqnarray}
v\left(x,y\right)=\left\{
\begin{array}{ccc}
-\epsilon_{Fn} & , & x'<-d\tfrac{\epsilon_{Fn}}{\epsilon_{Fn}+\epsilon_{Fp}}\\
-\epsilon_{Fn} + \tfrac{\epsilon_{Fn}+\epsilon_{Fp}}{d}\left(x'+d\tfrac{\epsilon_{Fn}}{\epsilon_{Fn}+\epsilon_{Fp}}\right) & , & -d\tfrac{\epsilon_{Fn}}{\epsilon_{Fn}+\epsilon_{Fp}}<x'<d\tfrac{\epsilon_{Fp}}{\epsilon_{Fn}+\epsilon_{Fp}}\\
\epsilon_{Fp} & , & x'>d\tfrac{\epsilon_{Fp}}{\epsilon_{Fn}+\epsilon_{Fp}}
\end{array}\right.
\label{onsiteD}
\end{eqnarray}
\end{widetext}
where $x'=x+I_{pn}(y)$ and $\epsilon_{Fn}=\left|\epsilon_{F}-\epsilon^{0}\right|$, $\epsilon^{0}$ being the Dirac point energy. $\epsilon_{Fp}$ is defined similarly. We had assumed that the pn junction is linearly graded, where the spatial extent of n-p transition (known as the ``depletion width'') is denoted by $d$. $I_{pn}$ is a one dimensional roughness profile generated using the above procedure. For $t_{ij}$ (in eV), we have,
\begin{eqnarray}
\left|t(x,y)\right|=\left\{
\begin{array}{ccc}
3 & , & I_{b}(x)<y<W+I_{t}(x)\\
0 & , & \text{otherwise}
\end{array}\right.
\label{couplingD}
\end{eqnarray}
$I_{b/t}$ is a one dimensional roughness profile describing the line edge roughness for the bottom and top edges respectively.

As an illustration, we modeled the magneto-transport across an armchair graphene pn junction with/without ID in a magnetic field of $B=10T$ with energies $\epsilon_{Fn}=0.1eV$ and $\epsilon_{Fp}=0.15eV$. Fig. \ref{currentplots} plots the longitudinal current component $j_{x}(x,y)$ of this device for different device width. The so-called ``snake states'' (i.e. current density oscillating back and forth the n/p regions) propagating along the pn interface can be observed \cite{beenakker08,pereira07,lukose07}, which remains prominent even in the presence of ID. The snake states terminate when the pn interface meets the top edge, where a choice between the paths leading to the n or p medium must be made. The ribbon's width, which also determines the valley isospins of the first LL along each edge, play a pertinent role in deciding which path is taken, as will be elaborated upon in the following section.

\section{\label{sec:level3}Valley isospins on the ballistic quantum Hall plateaus}

In this section, we examine the conductance across a pn junction when there is no disorder i.e. the ballistic limit. The ribbon's chirality plays an important role in determining the junction's conductance. Fig. \ref{IRorER}a depicts the notation on ribbon's chirality used in this work. In this limit, it was shown that conductance plateaus $\tilde{\sigma}_{pn}$ could emerged due to the valley isospins \cite{tworzydlo07} and wavefunction parity \cite{akhmerov08} of the ground state Landau level. The former effect concerns the armchair edge ribbon, while the latter for zigzag edge. An excellent review on this subject has been written by Beenakker \cite{beenakker08}, where we will highlight and expand on some of the key results in the remainder of this section. 

\subsection{\label{sec:level3a}Valley isospins along the edges: \\Definitions and conventions}
We write the Dirac equation for graphene as,
\begin{eqnarray}
H\Psi=\left[
\begin{array}{cc}
v_{f}\vec{p}\cdot\vec{\sigma} & 0 \\
0 & v_{f}\vec{p}\cdot\vec{\sigma}
\end{array}\right]\Psi
\label{diraceq}
\end{eqnarray}
where $\Psi=(\psi_{A},\psi_{B},-\tilde{\psi}_{B},\tilde{\psi}_{A})$ and $\psi$($\tilde{\psi}$) for the $\vec{K}$($\vec{K}'$) valley wavefunction. We are interested in $\Psi$ along the ribon's edges. It is a convenient convention to write $\Psi$ along the edges in the following form,
\begin{eqnarray}
\Psi=\left(\vec{v}\cdot\vec{\tau}\right)\otimes\left(\vec{n}\cdot\vec{\sigma}\right)\Psi
\label{bouneq}
\end{eqnarray}
where $\vec{v}$ is the $\bold{edge}$ $\bold{valley}$ $\bold{isospin}$ (for $\Psi$ along the edges) and $\vec{\tau}$ is just the Pauli matrices for the isospin part. $\vec{n}$ depends on the edge type i.e. $\vec{n}=(0,0,1)$ for zigzag and $\vec{n}=(\pm 1,0,0)$ for bottom/top edges of armchair ribbons \cite{beenakker08}. One can show that Eq. \ref{bouneq} effectively expressed the boundary conditions of the edges . For an armchair ribbon, it can be shown that the isospin along the top/bottom edges ($\vec{v}_{T}$ and $\vec{v}_{B}$) obeys the following,
\begin{eqnarray}
\vec{v}_{T}\cdot\vec{v}_{B}=cos\left(\Delta W +\pi\right)\equiv cos\theta
\label{iospin}
\end{eqnarray}
where $\Delta=4\pi/3a$ and $a$ is the lattice constant of graphene. From Fig. \ref{IRorER}, we have $W=a(l+\tfrac{1}{2})$, where $l$ is the number of carbon layers. See Appendix A for the detail algebra.

Next, one makes the assumption that the ground state LL's wavefunction, denoted by $\left|0\right\rangle$, could be approximated by the edge wavefunctions \cite{tworzydlo07}. This allows one to write the wavefunction overlap between the ground state LL's wavefunction along the top/bottom edges,
\begin{eqnarray}
\left\langle 0_{T}\right|\left.0_{B}\right\rangle&\approx&\left\langle \Psi_{T}\right|\left.\Psi_{B}\right\rangle\\
\nonumber
&=& \left(\left|a\right|^{2}+\left|b\right|^{2}\right)\left(1+\vec{v}_{T}\cdot\vec{v}_{B}+\vec{v}_{T}\times\vec{v}_{B}\right)
\label{overlapppp}
\end{eqnarray}
where we denote $(\psi_{A},\psi_{B})=(a,b)$ and $(\tilde{\psi}_{A},\tilde{\psi}_{B})$ is obtained through Eq. \ref{bouneq}. By using the fact that $|a|^{2}=|b|^{2}=\tfrac{1}{4}$, we finally arrive at,
\begin{eqnarray}
\left|\left\langle 0_{T}\right|\left.0_{B}\right\rangle\right|^{2}\approx\tfrac{1}{2}\left(1+cos\theta\right)
\label{overlapppp2}
\end{eqnarray}
Eq. \ref{overlapppp2} is a simple result \cite{tworzydlo07} that we will employ in the remaining of this section to draw some simple conclusions about the magnetotransport properties in a graphene pn junction. 

\begin{figure}[t]
\centering
\scalebox{0.53}[0.53]{\includegraphics*[viewport=170 200 620 490]{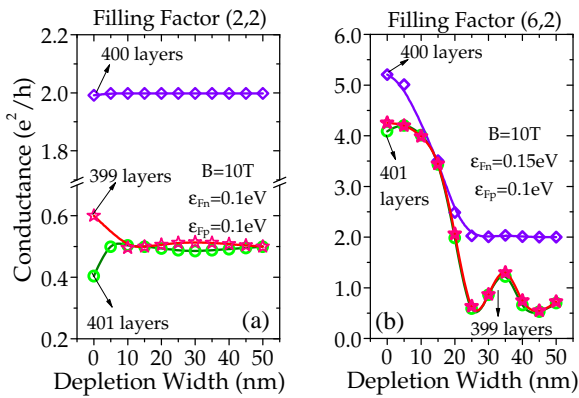}}
\caption{\footnotesize $\bold{(a)}$ Junction conductance as a function of depletion width in the clean limit i.e. no disorder, for armchair ribbons of different widths for the case of filling factor $(2,2)$. $\bold{(b)}$ Same as (a) excepts for filling factor of $(6,2)$.}
\label{armchairclean}
\end{figure}

\subsection{\label{sec:level3a}Valley isospin in armchair ribbon}

Fig. \ref{armchairclean} shows the ballistic conductance of armchair edge type ribbons as a function of depletion width. The n/p regions are biased at $\epsilon_{Fn/Fp}$ respectively, and the built-in potential (assumed to be linearly graded across the junction) is given by $\epsilon_{Fn}+\epsilon_{Fp}$. Fig. \ref{armchairclean}a plots the conductance for biasing conditions corresponding to the Landau filling combinations of $(\nu_{n},\nu_{p})=(2,2)$. Ribbons with different number of carbon layers along the width are considered, where the inter-layer separation is $\sqrt{3}L$ ($L$ being the carbon-carbon bond length). These ribbons exhibit conductance plateaus of $\tfrac{1}{2}$ and $2$ at sufficiently large depletion width of $>25nm$, where typical length scale of depletion width in experiments employing top/bottom gating scheme are usually several times larger than $25nm$ \cite{low09tt}. These plateaus emerge as long as the depletion width is sufficiently large, irregardless of the filling factor combinations. 

Fig. \ref{armchairclean}b plots the case when $(\nu_{n},\nu_{p})=(6,2)$, and we had also checked that these ballistic plateaus remain intact when $(\nu_{n},\nu_{p})=(6,6)$. It is observed that increasing depletion width filters off the higher Landau levels, such that only the zeroth mode Landau edge states conduct through the junction. This is reminiscent of the more well-known filtering action of off-normal transverse modes by a pn junction in the zero magnetic field case \cite{cheianov06,low09tt}, although the physics in this context is completely different. This might find applications in devices that use the Landau levels as an information bit \cite{beenakker06}. However, pn interface disorder would negate such filtering action, to be discussed in Sec. \ref{sec:level4}.

\begin{figure}[t]
\centering
\scalebox{0.95}[0.95]{\includegraphics*[viewport=180 330 620 620]{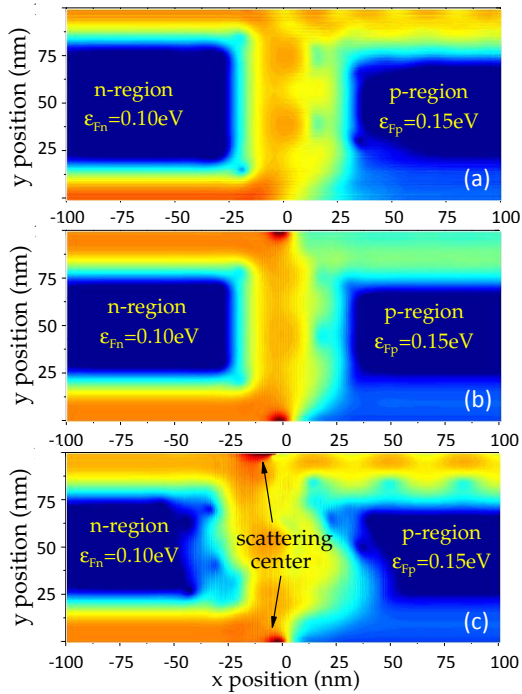}}
\caption{\footnotesize Intensity plot of the non-equilibrium electron density, $log_{10}(n)$, for various ribbons with width $\approx100nm$. We plotted for the case $\bold{(a)}$ armchair ribbon (no disorder) with $401$ carbon layers along the width, $\bold{(b)}$ zigzag ribbon (no disorder) and $\bold{(c)}$ zigzag ribbon (with interface disorder). The magnetic field is assumed to be $10T$, while the depletion width $25nm$. }
\label{densityplot}
\end{figure}

Tworzydlo and co-workers \cite{tworzydlo07} attributed the origin of the ballistic plateaus to the different valley isospins of the $0^{th}$ LL at the two edges of the ribbon. The valley isospins can be determined from their respective boundary conditions \cite{beenakker08}. When the number of layers satisfy the condition $3M+1$, where $M$ is an integer, it would exhibit conductance plateaus of $2$. Note that this is also the same criterion for obtaining a metallic armchair ribbon. This device is illustrated in Fig. \ref{currentplots}a and c for semiconducting and metallic type armchair ribbon respectively. We define the following scattering states for the $0^{th}$ LL; $\left|0_{nB}\right\rangle$ and $\left|0_{nT}\right\rangle$ for the incoming and reflected states, where the subscript n/p and T/B denotes the electron/hole mediums and top/bottom edges respectively. We can compute the reflection coefficient by following the prescription in \cite{beenakker08},
\begin{eqnarray}
r=\left\langle 0_{nT}\right|S\left|0_{nB}\right\rangle=\left\{
\begin{array}{cc}
0  & \mbox{3M+1}\\
\frac{\sqrt{3}}{2}e^{i\eta}  & \mbox{otherwise}
\end{array}
\right.
\label{overlapac}
\end{eqnarray}
The scattering matrix $S$ describing the evolution of the incoming scattering state $\left|0_{nB}\right\rangle$ along the pn interface can be simply described by a unit matrix with a constant phase factor. The conductance plateaus $\tilde\sigma_{pn}$ is then given by \cite{tworzydlo07},
\begin{eqnarray}
\tilde\sigma_{pn}=2(1-\left|r\right|^{2}) =\left\{
\begin{array}{cc}
2  & \mbox{3M+1}\\
\tfrac{1}{2}  & \mbox{otherwise}
\end{array}
\right.
\label{sigpn}
\end{eqnarray}
This argument requires the assumption that the valley isospin obeys the orthogonality identity $\vec{v}_{nT}\cdot\vec{v}_{pT}=\vec{v}_{nB}\cdot\vec{v}_{pB}=0$ for current conservation to hold \footnote{The sum of transmission and reflection probability must be unity. In other words, we should have $1-\left|\left\langle 0_{nB}\right|\left.0_{nT}\right\rangle\right|^{2}=\left|\left\langle 0_{nB}\right|\left.0_{pT}\right\rangle\right|^{2}$. This only holds true if $\vec{v}_{nT}\cdot\vec{v}_{pT}=\vec{v}_{nB}\cdot\vec{v}_{pB}=0$.}. 

\begin{figure*}[t]
\scalebox{0.66}[0.66]{\includegraphics*[viewport=50 344 750 550]{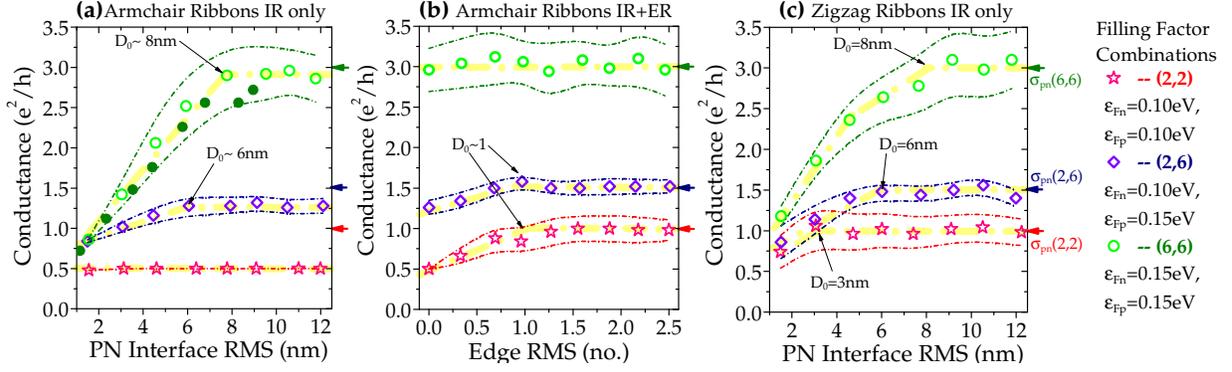}}
\caption{\footnotesize $\bold{(a)}$ Junction conductance as a function of IR disorder characterized by RMS for an armchair ribbon. All devices with open symbols have IR disorder of AL $\approx$ $7.5nm$, whereas those with solid symbols have AL $\approx$ $11nm$. No ER disorder and a ribbon width of $\approx$ $0.1 \mu m$, i.e. $401$ carbon layers, is assumed. Bias condition corresponding to filling factor combinations of $(2,2)$, $(2,6)$ and $(6,6)$ are plotted, see right inset for details. $\bold{(b)}$ Junction conductance as a function of ER disorder characterized by RMS (expressed in terms of number of layers) for an armchair ribbon. IR disorder assumed to have AL $\approx$ $7.5nm$ and RMS $\approx$ $11nm$. $\bold{(c)}$ Same as (a), except for zigzag ribbon counterpart. In these plots, each data point is an ensemble average over $100$ samples. The magnetic field is assumed to be $10T$ and the depletion width $25nm$. The dashed lines are plots of $\left\langle \sigma_{pn}\right\rangle\pm$var$\left(\sigma_{pn}\right)$. Thick dashed lines are drawn as guide to the eye so as to indicate the onset of conductance plateau.}
\label{disorder}
\end{figure*}

\subsection{\label{sec:level3b}Valley isospin in zigzag ribbon}
For zigzag ribbons, a similar width dependent effect can be observed, except that the conductance plateaus are $0$ and $2.0$ for zigzag and anti-zigzag ribbons respectively \cite{tworzydlo07}. These ballistic plateaus cannot be explained by the similar valley isospin argument as used for the armchair case. As first pointed out by Akhmerov and co-workers \cite{akhmerov08}, the reflected and transmitted edge states both reside on a valley different than the incident state i.e.
$\vec{v}_{nT}\cdot\vec{v}_{pT}=\vec{v}_{nB}\cdot\vec{v}_{pB}=1$. Therefore, current conservation entails an inherent intervalley scattering process.  Fig. \ref{densityplot} shows the intensity plot for the non-equilibrium electron density. Local peaks in the electron density can be observed at the positions where the pn interface and the ribbon edge meet. These are signatures of the intervalley scattering processes that have taken place. Heuristically speaking, one can view the propagating states along the pn interface as similar to that of an armchair edge, where the valley isospin is an equal weight superposition of the two valleys. The two times scattering process take the Landau state from one valley to a superposition and then finally to the other valley. In other words, the valley isospin information of the incident Landau edge state is intrinsically diluted after the first scattering process. On a related note, the local density of states in quantum Hall regime had been recently probed through scanning tunneling microscopy measurement \cite{hashimoto08}. It would be of fundamental importance to experimentally verify the existence of these localized intervalley scattering centers.

Akhmerov and co-workers \cite{akhmerov08} worked out the transmission/reflection coefficient for the edge states in the ballistic case (no magnetic field case) to be
\begin{eqnarray}
r=\left\langle 0_{nT}\right|S\left|0_{nB}\right\rangle=\left\{
\begin{array}{cc}
1  & \mbox{zigzag}\\
0  & \mbox{anti-zigzag}
\end{array}
\right.
\label{overlapzz}
\end{eqnarray}
where $S$ should embody the valley scattering processes. The conductance $\tilde\sigma_{pn}$ can then be computed in the same fashion as the armchair case. 
\begin{eqnarray}
\tilde\sigma_{pn}=\left\{
\begin{array}{cc}
0  & \mbox{zigzag}\\
2  & \mbox{anti-zigzag}
\end{array}
\right.
\label{sigpnzz}
\end{eqnarray}
By defining $\left\langle \tilde\sigma_{pn}\right\rangle$ to be the conductance plateaus averaged over the ribbon width, we obtained $\left\langle \tilde\sigma_{pn}\right\rangle_{ac}=\left\langle \tilde\sigma_{pn}\right\rangle_{zz}=1$.

\section{\label{sec:level4}Transition to Ohmic type quantum hall plateaus}

In this section, we examine the role of ID and ED on the quantum Hall plateau transition from ballistic-type to Ohmic-type i.e. $\tilde{\sigma}_{pn}$ to $\hat{\sigma}_{pn}$. Our numerical results show that this transition can be ID/ED-mediated depending on the filling factor combination $(\nu_{n},\nu_{p})$ and the ribbon type, as summarized in Fig. \ref{IRorER}b. The objective of this section is to present the argument as to why valley isospin dilution in general is necessary for the ballistic-Ohmic quantum Hall plateau transition, corroborated with numerical simulation results.

\subsection{\label{sec:level4a}Armchair edge ribbons}
Fig. \ref{disorder}a studies the junction conductance in the presence of pn interface disorder only, for an armchair ribbon of $401$ layers. The ballistic plateau for this device is $\tilde{\sigma}_{pn}=\tfrac{1}{2}$ since the number of carbon layers is $\neq 3M+1$. From a $(\nu_{n},\nu_{p})$-independent $\tilde{\sigma}_{pn}$ in the non-disordered limit, the junction conductance begins to adopt different $(\nu_{n},\nu_{p})$-dependent values as the ID RMS increases. The conductance saturates at large enough disorder strength. However, we observed that only the junction conductance for the $(6,6)$ case approaches the Ohmic values of $\hat{\sigma}_{pn}(6,6)=3$. In particular, the junction conductance for $(2,2)$ is extremely robust against IR disorder. The junction conductance for $(2,6)$ was enhanced with increasing IR disorder. However, it reaches a conductance plateau of only $\approx \tfrac{5}{4}$, lesser than the Ohmic value of $\hat{\sigma}_{pn}(2,6)=\tfrac{3}{2}$. In computing the conductance for a given RMS, we performed an ensemble averaging over $100$ devices with different roughness configurations. Fig. \ref{disorder}(b) plots the junction conductance as a function of edge disorder RMS, but with a fixed ID. Evidently, edge disorder with only a RMS of one carbon layer would suffices in inducing the plateau transition from $\tilde{\sigma}_{pn}$ to $\hat{\sigma}_{pn}$.  These observations suggest the following proposition, \itshape``In an armchair edge ribbon with filling factor combination of $(2,\nu_{p})$, the plateau transition from $\tilde{\sigma}_{pn}$ to $\hat{\sigma}_{pn}$ is both ID/ED-mediated. ''\normalfont

We consider a Chalker-Coddington \cite{chalker88} type argument in support of the above proposition. This model considers the following facts: (a) in the absence of time reversal symmetry, the electronic states exhibit only unidirectional transmission (b) the scattering wave function follows approximately the equipotential lines of the random potential, which is shown in Fig. \ref{currentplots}. Consider $(\nu_{n},\nu_{p})=(2,6)$, the scattering state for a particular spin along the pn interface can be expressed as,
\begin{eqnarray}
\left|\Psi_{i}\right\rangle=c_{0}\left|0_{nB}\right\rangle+c_{1}\left|0_{p}\right\rangle+c_{2}\left|1_{p}\right\rangle+c_{3}\left|1'_{p}\right\rangle
\label{scatstate}
\end{eqnarray}
where $\left|0_{p}\right\rangle$, $\left|1_{p}\right\rangle$ and $\left|1'_{p}\right\rangle$ are the ground and first excited states of the LL in the p medium respectively. We have $\vec{c}_{i}=(1,0,0,0)$ at the beginning of the pn interface. We can define a ``saddle point'' to be where two Landau modes $i$ and $j$ undergo mode mixing, characterized by a scattering matrix which evolves the scattering state $\left|\Psi\right\rangle$ in a unitary manner. The effective unitary matrix for four modes scattering can be parameterized as,
\begin{eqnarray}
S=\left[
\begin{array}{cccc}
c^{2} & sc & s^{2} & -sc\\
-sc & c^{2} & sc & s^{2}\\
s^{2} & -sc & c^{2} & sc\\
sc & s^{2} & -sc & c^{2}
\end{array}\right]
\label{smatrix}
\end{eqnarray}
where $s\equiv sin(\beta)$ and $c\equiv cos(\beta)$. As usual, the accompanied phase factors is implicit \cite{chalker88}. The parameter $\beta$ characterized the degree of mode-mixing i.e. $\beta=0,\tfrac{\pi}{4}$ denotes minimum/maximum mixing. Undergoing a sufficient amount of mode mixing processes $S$, the wavefunction at the end of the pn interface could be expressed as,
\begin{figure*}[t]
\scalebox{0.72}[0.72]{\includegraphics*[viewport=48 320 752 485]{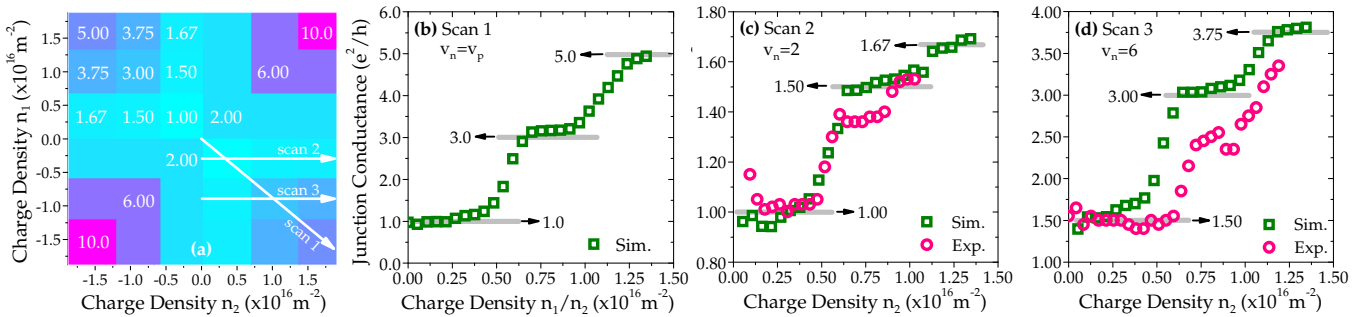}}
\caption{\footnotesize $\bold{(a)}$ depicts the theoretical Ohmic plateaus $\hat{\sigma}_{pn}(\nu_{n},\nu_{p})$ \cite{abanin07} as a function of $n_{1}/n_{2}$ where the different colors represent the filling factors. $\bold{(b)-(d)}$ plots the linescan for the following cases (i) $\nu_{n}=\nu_{p}$ (ii) $\nu_{n}=2$ and (iii) $\nu_{n}=6$ respectively, and compare the simulation results with experimental data. The experimental data are taken from \cite{williams07}, for a two terminal pn junction quantum Hall measurement at $B=4T$. The simulations are done at $B=10T$. In the experimental data, the gate oxide capacitance is used as a fitting parameter.}
\label{expcompare}
\end{figure*}
\begin{eqnarray}
\left|\Psi_{f}\right\rangle&=& S\left(\beta_{1}\right)S\left(\beta_{2}\right)S\left(\beta_{3}\right)\ldots\left|\Psi_{i}\right\rangle\\
\nonumber
&\approx& \tfrac{1}{2}e^{\phi_{0}}\left|0_{nB}\right\rangle+\tfrac{1}{2}e^{\phi_{1}}\left|0_{p}\right\rangle+\tfrac{1}{2}e^{\phi_{2}}\left|1_{p}\right\rangle+\tfrac{1}{2}e^{\phi_{3}}\left|1'_{p}\right\rangle
\label{scatstatef}
\end{eqnarray}
The final state is then said to have ``completely equilibrated''. The reflection probability can then be computed,
\begin{eqnarray}
\left|r\right|^{2}=\left|\left\langle 0_{nT}\right.\left|\Psi_{f}\right\rangle\right|^{2}=\tfrac{1}{4}\left|\left\langle 0_{nT}\right.\left|0_{nB}\right\rangle\right|^{2}\left|  e^{\phi_{0}}+e^{\phi_{1}}                \right|^{2}
\label{rr}
\end{eqnarray}
In arriving at the above result, we had make use of the orthogonality relation $\left\langle 0_{nT} \right.\left|1_{p}\right\rangle=0$. We also assumed that $\left|0_{p}\right\rangle$ retains the isospin information of the incident scattering state and therefore yielding us $\left\langle 0_{nT}\right|\left.0_{p}\right\rangle=\left\langle 0_{nT}\right|\left.0_{nB}\right\rangle$. By making use of the fact that the phase term averaging over a sufficiently large ensemble yields,
\begin{eqnarray}
\left\langle \left|  e^{\phi_{0}}+e^{\phi_{1}}+\ldots + e^{\phi_{n}}\right|^{2}\right\rangle_{ensemble}\approx n+1
\label{phaseran}
\end{eqnarray}
the junction conductance (including spin) at filling factor $(2,6)$ could then be expressed as, 
\begin{eqnarray}
\sigma_{pn}(2,6)\approx 2\left(1-\tfrac{1}{2}\left|\left\langle 0_{nT}\right.\left|0_{nB}\right\rangle\right|^{2}\right)
\label{tt}
\end{eqnarray}
For the case where the number of carbon layers of the armchair ribbon $\neq 3M+1$, Eq. \ref{tt} yields  $\sigma_{pn}(2,6)=\tfrac{5}{4}$.  Repeating the above analysis for filling factor $(2,2)$, we can show that $\sigma_{pn}(2,2)=\tilde{\sigma}_{pn}(2,2)$. The results from a simple Chalker-Coddington analysis are in excellent corroboration with what we obtained numerically from numerical calculations as shown in Fig. \ref{disorder}a. 

Remarkably, the above analysis predicts that the junction conductance for an armchair ribbon with $3M+1$ carbon layers will always be perfectly conducting i.e. $\sigma_{pn}(2,6)=\sigma_{pn}(2,2)=2$. Fig. \ref{currentplots}c-d depicts the spatial current density for this device at $(2,6)$ filling factor. Indeed, the transmission remains perfect in the presence of ID, despite the fact that mode mixing via ID has taken place. This fact unequivocally demonstrates that electron and hole Landau modes mixing via ID alone is not sufficient in achieving the ballistic-Ohmic plateau transition, at least for armchair ribbons. It is evident from Eq. \ref{tt} that the Ohmic result can be obtained if and only if $\left|\left\langle 0_{nT}\right.\left|0_{nB}\right\rangle\right|^{2}=\tfrac{1}{2}$. This is only possible if the isospin information on the T/B edges is completely diluted, e.g. via edge disorder. 

\subsection{\label{sec:level4b}Zigzag edge ribbons}
Fig. \ref{disorder}c study the junction conductance in the presence of pn interface disorder only, for a zigzag ribbon. The ballistic plateau for this device is $\tilde{\sigma}_{pn}=0$. In this case, the $\tilde{\sigma}_{pn}\rightarrow \hat{\sigma}_{pn}$ transition in the presence of ID is succinctly illustrated. Defining $D_{0}$ to be the disorder-related length scale where the junction conductance begins to plateau off, we obtained $D_{0}\approx 3,6,8nm$ for the $(2,2)$, $(2,6)$ and $(6,6)$ case respectively. Similar $D_{0}$ are observed for the armchair counterpart devices in Fig. \ref{disorder}a. This leads us to the following conclusion, \itshape``In a zigzag edge ribbon, ID alone is sufficient for the plateau transition $\tilde{\sigma}_{pn}$ to $\hat{\sigma}_{pn}$. In addition, it requires about the same disorder strength as its armchair counterpart to obtain the Ohmic plateaus, albeit ED is required for the latter.''\normalfont 

Repeating the Chalker-Coddington type argument, the reflection probability for the $(2,6)$ filling factor case can be written as,
\begin{eqnarray}
\left|r\right|^{2}=\tfrac{1}{4}\left|  0e^{\phi_{0}}+1e^{\phi_{1}}                \right|^{2}
\label{rr2}
\end{eqnarray}
by making use of Eq. \ref{overlapzz} and that $\left\langle 0_{nT}\right|\left.0_{p}\right\rangle=1-\left\langle 0_{nT}\right|\left.0_{nB}\right\rangle$. This then gives us a junction conductance of $\sigma_{pn}(2,6)=\tfrac{3}{2}=\hat{\sigma}_{pn}(2,6)$. Similarly, we can arrive at $\sigma_{pn}(2,2)=1=\hat{\sigma}_{pn}(2,2)$. Coincidentally, the signatures of the parity effect coming from the ground state LL is completely neutralized by the simple fact that $\left\langle 0_{nT}\right|\left.0_{p}\right\rangle=1-\left\langle 0_{nT}\right|\left.0_{nB}\right\rangle$. A similar physical picture applies for the anti-zigzag ribbon case.

\section{\label{sec:level6}Benchmarking with experiments}

In this section, we benchmark our numerical results against the two terminal quantum Hall measurement performed by Williams $et$ $al.$ \cite{williams07}. We should emphasize that bulk disorder was not included in present numerical simulation, therefore additional longitudinal resistance contributions which might exist in actual experiments are not captured \cite{williams09}. In the experiment, a top/bottom gate is employed to bias the two junctions, allowing control over the sign/magnitude of the charge density ($n_{1}/n_{2}$) residing in each junction. Fig. \ref{expcompare}a depicts the theoretical Ohmic plateaus $\hat{\sigma}_{pn}(\nu_{n},\nu_{p})$ \cite{abanin07} as a function of the electron density $n_{1}/n_{2}$ (the different colors represent the filling factors). In the numerical calculations, the electron density is obtained by taking the trace of the electron correlation function $G_{n}(\epsilon_{F})$, an energy resolved quantity. This is defined to be $G_{n}=G\Sigma_{in}G^{\dagger}$ where $\Sigma_{in}=-2$Im$(\Sigma_{s}+\Sigma_{d})$ for $T=0K$. Electron density in either the n/p medium can then be computed via the integral $n=\int \left\langle G_{n}\right\rangle d\epsilon$, where the averaging $\left\langle \ldots\right\rangle$ is performed over the spatial dimension. 

Fig. \ref{expcompare}b-d plots the linescan for the following cases (i) $\nu_{n}=\nu_{p}$ (ii) $\nu_{n}=2$ and (iii) $\nu_{n}=6$ respectively. In general, the numerical results show satisfactory agreement with the experiments. As previously addressed \cite{williams07}, the junction conductance with lower filling factors such as $\sigma_{pn}(2,2)$ and $\sigma_{pn}(2,6)$ plateau off at the expected Ohmic values of $1$ and $\tfrac{3}{2}$ respectively. However, higher plateaus such as $(6,6)$, $(6,10)$ could not be observed experimentally. This suggests that the interface disorder in the experiment is smaller than that necessary for complete Landau mode mixing of the higher plateaus. Fig. \ref{disorder}a indicates that ID with RMS and AL of $\approx 10nm$ should be sufficient. A more recent experiment by Lohmann $et$ $al.$ \cite{lohmann09} which employed chemical doping methods to create pn junctions should exhibits a larger ID. Although higher plateau measurements were not done in their experiments, their lowest plateau $\sigma_{pn}(2,2)$ exhibits a more precise plateau than that reported in \cite{williams07}. However, the improved precision in measurement is also a direct result of four terminal measurement \cite{lohmann09}. 

The general characteristics of the conductance during plateau transition also agree qualitatively with that of experiments i.e. the decreasing slope $\partial \sigma_{pn}/\partial n$ for the higher plateaus transitions. This is attributed to the smaller peak in the density of states (of the so-called extended states) of the higher LL \cite{streda82}. Another general remark can be made about the conductance plateau. Theoretically, the inter-LL energy spacing decreases with higher LL in a manner that is proportional to $\sqrt{n}-\sqrt{n-1}$, $n$ being the LL's index. However, the inter-LL spacing as function of charge density $n_{1/2}$ (inter-LL spacing herein denoted as $\delta_{n}$) is approximately equi-distant as shown in Fig. \ref{expcompare}a. The conductance plateau width for the filling factor $\nu_{n}=\nu_{p}$ could then be expressed as $\delta_{n}-\eta/(\partial \sigma_{pn}/\partial n)$, where $\eta=2e^{2}/h$. Since $\partial \sigma_{pn}/\partial n$ decreases with increasing $\nu$, the plateau width has to also decrease accordingly. This renders the observation of conductance plateau at higher LL more challenging in experiments. Nevertheless, our study suggests that the observations of junction plateaus up to filling factor of $6$ might be possible.

\section{\label{sec:level7}Summary }

In summary, we conducted a systematic study of graphene pn junction conductance in the quantum Hall regime. Often, the disorder in magneto-transport calculations are modeled in an implicit manner through a random on-site energy fluctuation. In this work, we had undertaken the effort to explicitly modeled the various disorder in order to uncover the underlying physical mechanisms played by pn interface and edge disorders on the ballistic-Ohmic quantum Hall plateau transition. We found that the former mechanism equilibrates the electron/hole Landau modes along the interface while the latter dilute the isospin information of the ground state LL in an armchair edge ribbon. For a given Landau filling factor combination, $(\nu_{n},\nu_{p})$, we found that both zigzag and armchair ribbons require about the same pn interface disorder strength to recover the Ohmic plateaus, albeit edge disorder is required for the latter. From our numerical calculations, we found that pn interface disorder with a root-mean-square roughness of $10nm$ is sufficient in achieving complete mixing of the electron/hole Landau modes. However, the sloppiness in the quantum Hall plateau transition induced by the pn interface disorder and the accompanied decreasing plateau widths with increasing filling factor makes it challenging to observe these higher plateaus experimentally.

From a theoretical standpoint, we argued that the mixing of electron/hole Landau modes along the interface alone does not guarantee the recovery of the Ohmic plateaus. We extended the valley isospin argument proposed by Tworzydlo and co-workers \cite{tworzydlo07} with a Chalker-Coddington \cite{chalker88} type argument to highlight this point, and corroborated the calculations with numerical simulations. It is demonstrated numerically that both interface and edge roughness (or intervalley scattering) are in general necessary for the crossover between the two theoretical limits i.e. ballistic and Ohmic. Last, but not least, this work underscores the importance of including both interface and edge disorder in the modeling of quantum Hall transport phenomena, especially when interpreting experimental data.

$\bold{Acknowledgement}$ We gratefully acknowledge support of the Nanoelectronic Research Initiative through the Institute for Discovery and Exploration (INDEX) and the generous computational support from Network for Computational Nanotechnology. T.L would like to thank M. Lundstrom and K. Wakabayashi for useful suggestions and discussions. 

\appendix

\section{\label{sec:appen} Boundary conditions and edge isospins for armchair ribbons}
We consider an armchair ribbon where the two edges are at $y=y_{T}$ (top) and $y=y_{B}$ (bottom). Along $y=y_{T}$, the wavefunction $\Psi=(\psi_{A},\psi_{B},-\tilde{\psi}_{B},\tilde{\psi}_{A})$ must satisfy the boundary conditions \cite{brey06},
\begin{eqnarray}
\psi_{A}+\tilde{\psi}_{A}e^{-i\Delta y_{T}}=0\\
\psi_{B}+\tilde{\psi}_{B}e^{-i\Delta y_{T}}=0
\label{bouneqtop}
\end{eqnarray}
where $\Delta=4\pi/3a$ and $a$ is the lattice constant of graphene. We can rewrite Eq. \ref{bouneqtop} in the form $\Psi=M\Psi$, where,
\begin{eqnarray}
M=\left[
\begin{array}{cc}
0 & -e^{-i\Delta y_{T}}\\
-e^{i\Delta y_{T}} & 0 
\end{array}\right]\otimes\left[
\begin{array}{cc}
0 & 1\\
1 & 0 
\end{array}\right]
\end{eqnarray}
With some matrix algebra we can show that,
\begin{eqnarray}
M=\left(\vec{v}_{T}\cdot\vec{\tau}\right)\otimes\left(\vec{n}_{T}\cdot\vec{\sigma}\right)
\end{eqnarray}
by defining $\vec{v}_{T}$=$\left(cos(\Delta y_{T}),sin(\Delta y_{T}),0\right)$ and $\vec{n}_{T}$=$(-1,0,0)$. Repeating this procedure for $y_{B}$, we require $\vec{v}_{B}$=$\left(-cos(\Delta y_{B}),-sin(\Delta y_{B}),0\right)$ and $\vec{n}_{B}$=$(1,0,0)$. It is straightforward to see that,
\begin{eqnarray}
\vec{v}_{T}\cdot\vec{v}_{B}=cos\left(\Delta W +\pi\right)\equiv cos(\theta)
\end{eqnarray}
where $W=y_{T}-y_{B}$.


\end{document}